\documentclass[a4paper,12pt,preprint,aps,floatfix]{revtex4}
\usepackage{longtable}
\usepackage{natbib}
\usepackage{graphicx}     
\usepackage{lscape}
\usepackage{color}

\newcommand{\etal}{{\it et al.} }
\newcommand{\ai}{{\it ab initio }}

\newcommand{\cm}{cm$^{-1}$}

\newcommand{\hp}{H$_3^+$}
\newcommand{\htwo}{H$_2$}

\def\a0{{$a_{\rm 0}$}}

\begin{document}
\hspace{3cm}
\title{ QED correction for    H$_3^+$}


\author{
 Lorenzo Lodi,$^{1}$   Oleg L. Polyansky$^{1,2,*}$, Jonathan Tennyson,$^{1}$ 
Alexander Alijah$^{3}$ and
Nikolai F. Zobov$^{2}$ } 
\affiliation{$^{1}$Department of Physics and Astronomy, University College London, Gower Street,  London WC1E~6BT, UK.}
\affiliation{$^{2}$Institute of Applied Physics, Russian Academy of Sciences,
Ulyanov Street 46, Nizhny Novgorod, Russia 603950.}
\affiliation{$^{3}$Groupe de Spectrom\'etrie Mol\'eculaire et Atmosph\'erique,
GSMA, UMR CNRS 6089, Universit\'e de Reims Champagne-Ardenne, France}
\email{o.polyansky@ucl.ac.uk}
\label{firstpage}

\date{\today}

\begin{abstract}
A quantum electrodynamics (QED) correction surface 
for the simplest polyatomic and polyelectronic system
\hp{} is computed using an approximate procedure.
This surface is used to calculate the shifts 
to vibration-rotation energy levels due to QED;
such shifts have a magnitude of up to 0.25~\cm{}
for vibrational levels up to 15~000~\cm{} and are
expected to have an accuracy of about 0.02~\cm.
Combining the new \hp\ QED correction surface with existing
highly accurate Born-Oppenheimer (BO), relativistic
and adiabatic components 
suggests that deviations of the resulting {\it ab
initio} energy levels from observed ones  
are largely due to non-adiabatic effects.

\pacs{Valid PACS appear here}

\end{abstract}

\maketitle

\section{Introduction}
\emph{Ab initio} studies of diatomic and triatomic systems containing 
less than ten electrons are nowadays able to produce rotation-vibrational
energy levels with better than spectroscopic accuracy, i.e. with errors
of less than 1~\cm. To improve on this  accuracy one
needs to account for several small effects which are routinely neglected,
including electronic relativistic and adiabatic corrections, as well as --- 
most notably for this work --- non-adiabatic effects and
corrections due to quantum electrodynamics (QED).
General discussions of relativistic and QED effects in molecular physics and
quantum chemistry can be found in several recent
reviews \cite{Pyykko2012,Liu2013,Autschbach2012,Liu2012,Kutzelnigg2012,Ilias2010,anatomy} and
textbooks \cite{Reiher.book,Dyall.book}.
In this study we follow the convention of calling `relativistic effects'
corrections to the non-relativistic Schr{\"o}dinger equation
of second order in the fine-structure constant $\alpha$
(i.e., all effects correctly described by the many-electron no-pair
Dirac-Coulomb-Breit equation), while so-called radiative
corrections due to the quantization of the electromagnetic field
and appearing in higher powers of $\alpha$ are referred
to as QED effects.

The hydrogen molecular ion H$_2^+$ is the simplest physical
system with a rotational-vibrational spectrum and serves as
an important benchmark.
Rotational-vibrational energy levels for H$_2^+$
were notably presented by Moss \cite{Moss1999}
with an estimated accuracy of $10^{-4}$~\cm{}
and included non-adiabatic, relativistic
as well as leading QED corrections.
More recent studies have considerably improved the achievable 
accuracy and, 
for selected rotation-vibrational transitions, QED corrections up to $\alpha^5$
have been computed \cite{Korobov2006,Korobov2008,Zhong2012} leading
to uncertainties of about $2\times 10^{-6}$~\cm.

Next in terms of size and complexity is the hydrogen molecule \htwo,
for which an accuracy of  $10^{-4}$  \cm\ has recently been achieved
{\it ab initio} \cite{KomasaJCTC2011, Ubachs,Ubachs1}
by careful inclusion of non-adiabatic corrections and of QED corrections
to order $\alpha^4$.
Studies of H$_2^+$ and H$_2$ represent the current state-of-the-art
for calculations of molecular rotational-vibrational energy levels;
for larger systems the achievable accuracy is considerably lower.

In particular, for \hp\ the highest accuracy achieved so far is 0.10~\cm\ for
all known energies up to 17~000~\cm\ \cite{jt512}, which is therefore
several orders of magnitude worse than for H$_2^+$ and H$_2$.
Higher accuracy energy levels are necessary for proper analysis of \hp\
experimental spectra. More specifically, about 30 years ago Carrington and
co-workers \cite{CB82,CK84,cm89} measured very dense near-dissociation
spectra of \hp\ and its isotopologues with
an average line spacing of less than 0.01 \cm; these spectra,
which remain unassigned and substantially uninterpreted \cite{jt157},
clearly require very high accuracy to be analysed from theoretical calculation.

Another source of motivation is provided  by
the recent studies by Wu {\it et al} \cite{13WuLiLi.H3+}
and Hodges  {\it et al} \cite{13HoPeJe.H3+}, who have concentrated
on high-precision and high-accuracy frequency measurements on
the \hp\ $\nu_2$ fundamental band.
Measurements were made by both groups at the sub-MHz ($3 \times 10^{-5}$ \cm)
level but currently do not agree with each other within the claimed uncertainties.

The assigned \hp\ experimental data has recently been the subject of an
analysis using the MARVEL procedure \cite{jt412},
producing a comprehensive set of rotation-vibration energy levels
\cite{H3pMARVEL,H3pMARVEL1} which we use for comparison throughout this study.

Given the present experimental situation it is therefore very desirable to
improve the accuracy of theoretical \hp\ energy levels beyond the 0.1~\cm\ level.
The main non-relativistic, clamped nuclei Born-Oppenheimer (BO)
potential energy surface (PES) from  Pavanello  {\it et al}  \cite{jt512,jt526} 
and the associated
relativistic and adiabatic surfaces, all of which we use in this work,
are probably sufficiently well-determined to predict energy levels
with an accuracy of about $10^{-2}$ \cm\  for low-lying
levels up to about 15,000~\cm. 
There are currently two factors limiting the accuracy in \hp\ to the 0.1~\cm{} level,
namely a proper treatment
of, \emph{i)}, non-adiabatic and, \emph{ii)}, QED effects.

Non-adiabatic effects in \hp\ and its isotopologues
are known to affect line positions by up to
                             1.0~\cm{} \cite{jt236} and therefore
must be accounted for accurately.
Polyansky and Tennyson (PT) \cite{jt236} introduced a simple
model based on the use of fixed, effective vibrational
and rotational masses taken from  Moss's \cite{Moss1993} studies on H$_2^+$;
PT were able to improve the accuracy of
calculations from 1~\cm\ to 0.1~\cm.
Further improvements require more sophisticated treatments of non-adiabatic effects;
a step in this direction has been made by  Diniz {\it et al}
\cite{jt566}, who obtained non-adiabatic 
rotational-vibrational energies for the $\nu_2$ band with an 
accuracy of      0.01    ~\cm\ but did not consider higher vibrational states.

The second factor limiting the final accuracy of \hp\ energy levels are QED effects.
As discussed above, QED effects have been computed accurately for
H$_2^+$ \cite{Moss1993} and H$_2$ \cite{KomasaJCTC2011,Ubachs,Ubachs1}
and have an effect in the region 0.1---0.2~\cm\ on the corresponding rotation-vibration energy levels. 
In the case of \hp, QED effects have so far been  entirely neglected
but must clearly be taken into account to achieve
accuracies better than 0.1~\cm.

Pyykk{\"o} \emph{et al.} \cite{jt236} suggested a simple scheme for
describing leading QED effects in molecules (see section \ref{section.QED} for details).
This scheme has been already applied to the water molecule \cite{jt236,jt309}
--- for which QED corrections 
are of the order of 1~\cm --- and was instrumental in
recent studies achieving an accuracy of 0.1~\cm\ for levels
up to 15~000~\cm\ \cite{jt550} and of 1~\cm\ for the dissociation energy \cite{jt549}.
In this study we use the model of Pyykk{\"o} \emph{et al.} \cite{jt236}
to provide a QED correction surface for \hp.
This correction energy surface, when combined with the existing
non-relativistic, relativistic and adiabatic
surfaces from previous studies \cite{jt512,jt526} and with a
future, accurate treatment of non-adiabatic effects is expected
to provide rotation-vibration energy levels with a typical accuracy of
0.01 \cm.

The paper is organised as follows.
Section \ref{sec.basis-set} presents a comparison of the
Born-Oppenheimer PES computed using explicitly
correlated Gaussians \cite{jt512,jt526} and surfaces computed using
standard quantum chemistry methods based on full configuration
interaction (FCI) and Gaussian basis sets.  We  show that
available basis sets provide an accuracy between 0.1 \cm\ and 1 \cm\ for rotation-vibration
energy levels.  Section III compares results of accurate QED calculations
for \htwo\ \cite{KomasaJCTC2011,Ubachs,Ubachs1} with our calculations
using the approximate method of Pyykk{\"o} et al. \cite{jt236}.
QED corrections for \hp\ using the  same methodology are presented.
Section  IV presents results of nuclear motion
calculations using a BO PES, relativistic and adiabatic corrections
\cite{jt512,jt526} and our QED correction surface. Nuclear motion
calculations are given both without non-adiabatic corrections and 
with a simple non-adiabatic treatments based either on the Polyansky-Tennyson
(PT) model  \cite{jt236} or on the model by Diniz {\it et al}
\cite{jt566}.  Analysis of the residual deviations between
theory and experiment is given. Section V presents a final
discussion and conclusions.

\section{Errors due to basis set incompleteness for \htwo{} and \hp}\label{sec.basis-set}
Before discussing QED corrections we briefly discuss errors in
vibrational energy levels computed from non-relativistic BO energy
surfaces obtained using standard quantum chemistry methods.  We find
this discussion appropriate because practical application of the
method of Pyykk{\"o} \emph{et al.} \cite{jt236} for QED correction
also relies on standard electronic structure methods.  All
calculations used the electronic structure program Molpro
\cite{MOLPRO2012} using the CISD (configuration interaction single and
doubles) method; because \htwo{} and \hp\ are two-electron systems
CISD for these systems is equivalent to full CI (FCI); this means that
electron correlation is accounted for exactly and the error in
non-relativistic energies is entirely due to basis set incompleteness.
In all calculations we used the aug-cc-pV$n$Z correlation-consistent
family of basis sets introduced by Dunning \cite{Dunning1989} with $n
=$ D, T, Q, 5 and 6; these will be referred to by the shorthand
notation a$n$z. Two-term basis-set extrapolated values used the
extrapolation formula $E_n = E_\infty + A / n^4$ and are denoted
a$[n,m]$z; as discussed below, this extrapolation form was used
because it gives the best agreement with very accurate reference
results for H$_2$.
For comparison, we also include results obtained using
explicitly-correlated methods of the F12 family \cite{Shiozaki2013,Hatting2012,Kong2012,Ten-Ho2011};
in particular, we used the CISD-F12 code available in Molpro \cite{Shiozaki2011}. 

We did not include H$_2^+$ in this comparison because it is a one-electron system
and it is well known \cite{Kong2012} that basis set incompleteness error
is dominated by the electron correlation part, so that basis set convergence
results for H$_2^+$ are not representative of many-electron systems.

\subsection{Non-relativistic surfaces}
Our Molpro-based results were compared with much more accurate
calculations performed using explicitly correlated exponentials \cite{Pachuz} (\htwo)
and explicitly correlated Gaussians (ECG) \cite{jt512} (\hp); these reference
values should provide clamped-nuclei Born-Oppenheimer energies with an accuracy
of at least $10^{-4}$~\cm\ for \htwo\ and $10^{-3}$~\cm\ for \hp\ and
will be referred to as `exact' below.
 
\begin{table}
\caption{Errors in H$_2$ vibrational ($J=0$) energy levels computed from FCI non-relativistic energy curves and various basis sets (see text for details). All values are in \cm. \label{table.H2}}
\mbox{}\\
\begin{tabular}{r@{\hskip 0.15in}r@{\hskip 0.15in}r@{\hskip 0.15in}r@{\hskip 0.15in}r@{\hskip 0.15in}r@{\hskip 0.15in}r@{\hskip 0.15in}r @{\hskip 0.15in}}
\hline
\hline
    &    exact$^a$ & \multicolumn{6}{c}{exact - calculated} \\ \cline{3-8}
$v$ &                                       &   a4z      &    a5z    &   a6z     &  a[5,6]z$^b$&  a4z/F12 & a5z/F12 \\ 
\hline
 0  &        0.00                           &   0.00     &   0.00    &  0.00     &   0.00      &  0.00 &  0.00   \\ 
 1  &    4~163.40                           &   4.43     &   1.33    &  0.65     &   0.02      &  0.92 &  0.06   \\ 
 2  &    8~091.16                           &   8.68     &   2.64    &  1.36     &   0.17      &  1.82 &  0.08   \\ 
 3  &   11~788.14                           &   12.79    &   3.97    &  2.07     &   0.30      &  2.57 &  0.06   \\ 
 4  &   15~257.39                           &   16.82    &   5.41    &  2.83     &   0.43      &  3.15 &  0.02   \\ 
 5  &   18~499.88                           &   20.89    &   6.99    &  3.65     &   0.54      &  3.61 & -0.03   \\ 
 6  &   21~514.30                           &   25.16    &   8.77    &  4.57     &   0.65      &  3.97 & -0.11   \\ 
 7  &   24~296.64                           &   29.80    &   10.78   &  5.60     &   0.77      &  4.24 & -0.25   \\ 
 8  &   26~839.64                           &   35.01    &   13.07   &  6.79     &   0.94      &  4.46 & -0.42   \\ 
 9  &   29~131.99                           &   41.01    &   15.71   &  8.18     &   1.18      &  4.69 & -0.60   \\ 
 10 &   31~157.32                           &   48.08    &   18.78   &  9.82     &   1.48      &  5.02 & -0.76   \\ 
 11 &   32~892.55                           &   56.51    &   22.40   &  11.78    &   1.89      &  5.57 & -0.87   \\ 
 12 &   34~305.64                           &   66.62    &   26.76   &  14.17    &   2.44      &  6.39 & -0.89   \\ 
 13 &   35~352.20                           &   78.85    &   32.19   &  17.14    &   3.13      &  7.63 & -0.72   \\ 
 14 &   35~970.80                           &   94.16    &   39.21   &  20.97    &   4.00      &  9.81 & -0.12   \\ 
\mbox{}\\
\multicolumn{2}{r}{RMS$^c$=}                &   45.04    &  18.00    &    9.53   & 1.66         & 4.90 &   0.47\\
\hline
\hline
\end{tabular}
\mbox{}\\
\begin{flushleft}
$^a$ Using the very accurate BO potential energy points by Pachucki \cite{Pachucki2010}.\\     
$^b$ Using the extrapolation formula $E_n = E_\infty + A / n^4$.\\
$^c$ Root-mean-square deviation.
\end{flushleft}
\end{table}

\begin{figure}
\begin{center}
\includegraphics[angle=0, width=0.75\textwidth]{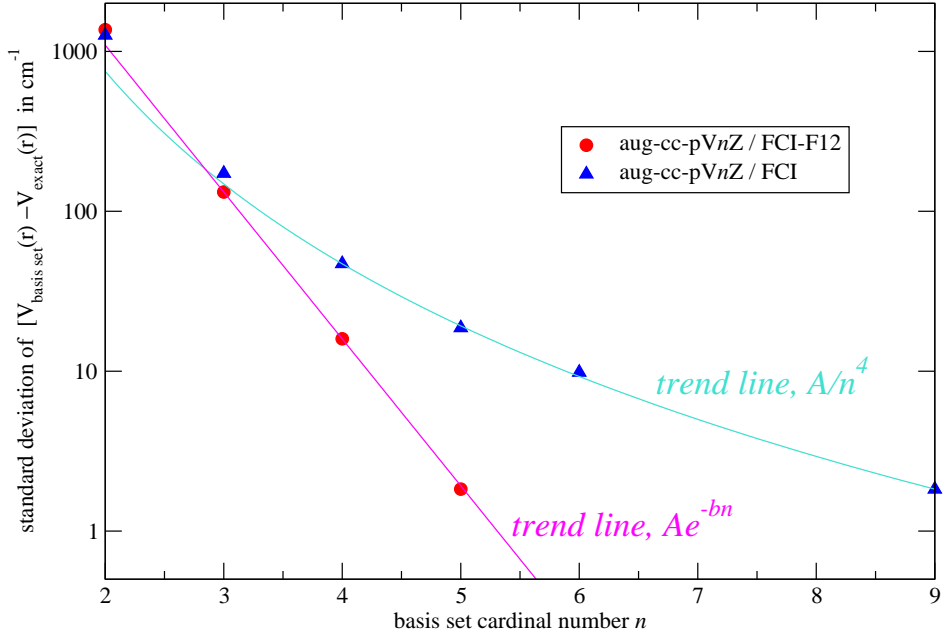}
\caption{Convergence speed of H$_2$ relative energies with various basis sets. The quantity plotted is the standard deviation of the difference between a potential curve
obtained with a given basis set and the virtually exact one from ref. \cite{Pachucki2010} in the range $r=0.75$~a$_0$ to $r=12$~a$_0$. The trend lines are fits to the last three points of each series.}
\label{fig.H2}
\end{center}
\end{figure}

Results for \htwo\ are collected in table~\ref{table.H2} and represented in figure~\ref{fig.H2}.
An analysis of the convergence pattern reveals that FCI errors 
decrease with the basis set cardinal number $n$ with an $n^{-4}$ dependence; for this reason
the basis set extrapolation formula $E_n = E_\infty + A / n^4$ works best for this system and was used
throughout. This observation is in line with several recent studies \cite{Peterson2012,Feller2011} which
show very good performance for the similar formula $E_n = E_\infty + A / (n+1/2)^4$ with respect
to other basis set extrapolation schemes.
As a result of this regular convergence behaviour extrapolated a[5,6]z
energy levels improve very significantly over the raw a6z values and have an accuracy
comparable with the expected one for the a9z basis set. In particular, the error of
a[5,6]z vibrational energy levels is very nearly linear up to $v=9$ and has an approximate
magnitude of $0.12~v$~\cm. As discussed in detail below, similar basis set errors are found for \hp .  

Explicitly-correlated methods of the F12 type do exceptionally well for H$_2$ and show
exponential convergence in terms of $n$ (see table \ref{table.H2} and fig.~\ref{fig.H2}); as a result
a5z/F12 energy levels are of overall higher quality than extrapolated a[56]z ones, especially
for energies above 20~000~\cm.
We also considered the basis sets of the cc-pV$n$Z-F12 family ($n=$ D, T and Q) 
\cite{Peterson2008,Yousaf2008} especially designed for F12 calculations; these basis sets too show exponential convergence and, moreover,
reduce errors with respect to the corresponding a$n$z basis set by a factor 7 for
a2z and by a factor 3 for a3z and a4z.


The first FCI calculations for \hp\  were performed in a
classic 1986 work by Meyer, Botschwina and Burton (MBB) \cite{MBB}; 
subsequent studies 
gradually increased the accuracy of the PES and extended its range.
Most of this work was performed \ai\ \cite{lf92,Cencek1998,jt247,Bachorz2009,jt512, VIE07:074309,Velilla2008}
but in a few cases the PES was improved by fitting to spectroscopic data \cite{jt141,jt168, jt202,jt512}.
These theoretical studies proved indispensable for the assignment of new observed lines of \hp,
see for example refs. \cite{jt83,jt102,jt193,jt413}.


We performed FCI calculations at the 69 geometries originally used by MBB  \cite{MBB}
for \hp\ using the same methodology described above for H$_2$; energies were fitted in a
standard way, following the procedure  described previously \cite{jt236}. These calculations are 
compared to the high-accuracy values computed by  Cencek \emph{et al.}
\cite{Cencek1998} instead of the more recent and accurate one by Pavanello \emph{et al.}
\cite{jt512,jt526} used elsewhere in this work because the latter were computed on a different grid.
The results of Cencek \emph{et al.} are sufficiently accurate for this purpose and will be labelled as `exact'
below.
The results of the vibrational $J=0$ energy levels are reported in table~\ref{table:h3p}.
Explicitly correlated F12 methods show improved convergence speed but not quite as fast as for H$_2$;
as a result extrapolated a[56]z and a5z-F12 energy levels have comparable accuracies (see table~\ref{table:h3p}).


\begin{table}
\caption{Errors in \hp\ vibrational band origins ($J=0$) energy levels computed from FCI non-relativistic energy curves and various basis sets (see text for details). All values are in \cm.   \label{table:h3p}}
\begin{tabular}{lr@{\hskip 0.20in} r@{\hskip 0.20in}r@{\hskip 0.20in}r@{\hskip 0.20in}r@{\hskip 0.20in}r@{\hskip 0.20in}r}
\hline
\hline
$(v_1,v_2^\ell)$&   exact$^a$ &     \multicolumn{4}{c}{exact - calculated}    \\ \cline{3-6}
                &             &     a5z    &    a6z & a5z/F12&    a[5,6]z$^b$   \\
\hline
  $(0, 1^1)$    &    2~521.51 &    0.74    &  0.47  &  0.17  &          0.11   \\
  $(1, 0^0)$    &    3~179.59 &    0.44    &  0.35  &  0.07  &          0.16   \\
  $(0, 2^0)$    &    4~778.34 &    1.58    &  0.94  &  0.32  &          0.12   \\
  $(0, 2^2)$    &    4~998.31 &    1.52    &  0.92  &  0.32  &          0.14   \\
  $(1, 1^1)$    &    5~555.42 &    1.17    &  0.80  &  0.23  &          0.26   \\
  $(2, 0^0)$    &    6~264.44 &    0.90    &  0.66  &  0.13  &          0.30   \\
  $(0, 3^1)$    &    7~006.10 &    2.42    &  1.40  &  0.46  &          0.42   \\
  $(0, 3^3)$    &    7~285.50 &    2.52    &  1.43  &  0.44  &          0.44   \\
  $(1, 2^0)$    &    7~770.20 &    2.87    &  1.18  &  0.33  &         -0.09   \\
  $(1, 2^2)$    &    7~870.84 &    2.04    &  1.26  &  0.35  &         -0.00   \\
  $(2, 1^1)$    &    8~489.38 &    1.70    &  1.11  &  0.25  &         -0.03   \\
  $(0, 4^0)$    &    9~001.04 &    3.32    &  1.87  &  0.58  &          0.62   \\
  $(0, 4^2)$    &    9~112.17 &    3.44    &  1.90  &  0.56  &          0.74   \\
  $(3, 0^0)$    &    9~254.77 &    1.43    &  0.97  &  0.16  &          0.17   \\
  $(1, 3^1)$    &    9~653.33 &    3.00    &  1.75  &  0.45  &         -0.51   \\
  $(1, 3^3)$    &    9~966.80 &    2.91    &  1.68  &  0.16  &         -0.96   \\
  $(0, 4^4)$    &    9~997.51 &    2.54    &  1.58  &  0.36  &         -0.96   \\
  $(2, 2^0)$    &   10~592.76 &    2.15    &  1.37  &  0.11  &         -2.22   \\
  $(2, 2^2)$    &   10~643.45 &    2.31    &  1.46  &  0.02  &         -2.41   \\
  $(0, 5^1)$    &   10~855.91 &    3.83    &  1.90  & -0.08  &         -0.18   \\
\mbox{}\\
\multicolumn{2}{r}{RMS$^c$=}  &    2.33    &  1.33  &  0.32  &          0.85   \\
\hline
\hline
\end{tabular}
\mbox{}\\
\begin{flushleft}
$^a$ Using the very accurate BO potential energy surface by Cencek \emph{et al} \cite{Cencek1998}. \\
$^b$ Using extrapolation formula $E_n = E_\infty + A / n^4$.\\
$^c$ Root-mean-square deviation in \cm.
\end{flushleft}
\end{table}


Our FCI-based energy levels for \htwo\ have a root-mean-square (RMS) deviation with respect to the exact reference values
of 1.66~\cm{} (extrapolated a[56]z energies)
or 0.47~\cm{} (a5z-F12 energies); the RMS errors for \hp\ in the energy range to 10~000~\cm{} are 0.85~\cm{} (a[56]z)
and 0.32~\cm{} (a5z-F12). We conclude that F12 methods at the a5z 
level are capable of providing energy levels accurate to better than 0.5~\cm, and a[56]z generally
to better than 1.5~\cm. Such calculations are therefore
a viable, good-quality alternative when explicitly-correlated Gaussian methods are too expensive.


\subsection{Relativistic surfaces}\label{sec.rel}
The most accurate relativistic corrections for \hp{} are those
by Bachorz \emph{et al} \cite{Bachorz2009} and were computed as
expectation value
(using a very accurate wave function based on explicitly correlated
Gaussians) of
the complete Breit-Pauli relativistic Hamiltonian \cite{Bethe-Salpeter},
i.e.
including mass-velocity, one- and two-electron Darwin contributions,
Breit retardation and spin-spin Fermi contact term.
The relativistic correction for \hp{} is overall very small, spanning
the range -4.3 to -1.8~\cm{} over all the geometries considered. 
As discussed below such a small contribution is due
to almost complete cancellation between the main contributions to
the overall relativistic correction.

We used Molpro \cite{MOLPRO2012} to compute relativistic
corrections
as expectation value of the mass-velocity (MV) and
one-electron Darwin (D1) operator using full-CI wave functions.
The Molpro-based aug-cc-pV6Z MVD1 energies are converged with respect to
basis set to about 0.05~\cm;
they typically agree with the more complete relativistic corrections by Bachorz to 0.15~\cm, which
can only be considered a moderate agreement considering the overall
smallness of the relativistic
correction. This should indicate that the contribution to relative energies of
terms neglected in the Molpro-based calculation (Breit, two-electron
Darwin and Fermi contact)
are non-negligible for very accurate work.
On the other hand this also
indicates that the two-electron QED correction (based on the
two-electron Darwin contribution) should be negligible, as it is expected
to be about 6 times smaller than the one-electron part.

It is worth performing a more detailed analysis of the MVD1 correction.
The MV term has an absolute magnitude of about -23~\cm, while the D1 term
of about +20~\cm; both contributions show a variation with geometry
spanning about 6~\cm.
However, the variation with geometry of MV and D1 are almost perfectly
anti-correlated
resulting in mutual cancellation when summed.
As a result of this cancellation the MVD1 contribution turns out to be
only slightly larger than the QED one (see section \ref{section.QED}).

The situation is somewhat similar for water (analysis performed for
energies up to 40~000 \cm) \cite{jt475}.
The MV term is in absolute terms (average value) -57~000~\cm\ with a
variation of 500~\cm, and D1 +45~000~\cm\ with a variation of 400~\cm.
The MVD1 term has a magnitude of -11~500~\cm{} with a variation of 140~\cm.
The QED correction for water is 1~000~\cm\
with a variation of 2~\cm.
So in the case of water there still is considerable cancellation, but
not as much as in \hp.
 
%

\section{Quantum electrodynamics corrections for  \htwo{} and \hp}\label{section.QED}
Pyykk{\"o} \emph{et al} \cite{jt265,Pyykko2012} proposed
making use of approximate proportionality formulae
between the leading QED corrections to order $\alpha^3$ (namely, the electron self-energy)
and the one- and two-electron Darwin corrections.
We neglect the two-electron contribution and compute the one-electron
Darwin term with Molpro and FCI wavefunctions; as discussed in section \ref{sec.rel} the two-electron
contribution is expected to be about a factor 6 smaller than the one-electron one.
Pyykk{\"o} \emph{et al}'s method
requires a scaling factor for which we use  0.04669,
as reported in table II of  Pyykk{\"o} \emph{et al} for all systems studied. 


QED corrections are known accurately both for H$_2^+$ \cite{Moss1993} and \htwo{} \cite{KomasaJCTC2011, Ubachs, Ubachs1};
we compare our scheme with these reference calculations in tables \ref{table.qed.H2plus} and \ref{table.qed.H2}.

The QED values differ on average from exact  ones  by less than 0.001 \cm{} for H$_2^+$ and less than 0.02 \cm{} for \htwo
 (see tables III and IV). 
Columns three and four of table~\ref{table.qed.H2} give the
relativistic and QED shifts in the energy levels of \htwo\ from the
exact calculations \cite{KomasaJCTC2011}.  Column 6 gives the
relativistic FCI a[5,6]z calculation of MVD1 using Molpro and the
column 7 gives the scaled by 0.04669 value of column 6, which gives
our approximate QED value. One can see that the exact shifts
differ from our approximate calculations by 0.02 \cm\ or
less for all except the highest, $v=14$ vibrational level.  We express
the substantiated hope here that the QED calculation for \hp, given
below, deviates from any future exact calculation by not much more
than this value.

Let us now consider our analogous QED calculations for \hp. The MVD1
calculations were also performed using Molpro and the a6z CBS basis
set.  However, our comparison of these calculations with one performed
using a aQz basis set showed rapid convergence of the relativistic
calculations with basis set, so in practice our aQz results could have
been also used. 
 Table~ \ref{table.qedMBB} gives values for the calculated QED 
corrections at all 69 MBB geometries. It can be seen the magnitude of the QED correction is small, less than 1~\cm\ everywhere,
but that it varies significantly with geometry and even changes sign.
We fitted the 69 QED points computed at the a6z level
to the functional form used if ref. \cite{jt512} to fit the relativistic
energies. 
The function contained 9 fitting parameters, polynomials up to degree 4 and
reproduced the \emph{ab initio} values with a root-mean-square deviation
of 3.3$\times$ 10$^{-3}$~\cm.

\begin{table}
\caption{QED corrections for $J=0$ vibrational levels of H$_2^+$.\label{table.qed.H2plus}}
\mbox{}\\
\begin{tabular}{r r @{\hskip 0.30in} r @{\hskip 0.25in}r r r @{\hskip 0.25in}r r r }
\hline
\hline
       &        BO$^a$&   \multicolumn{7}{c}{QED corrections}\\
\cline{3-9}
       &              &   \multicolumn{1}{c}{exact$^b$}  &  \multicolumn{3}{c}{this work} & \multicolumn{3}{c}{exact -- this work} \\
$v$    &              &                &     a4z      &      a5z     &    a6z   &     a4z      &      a5z      &    a6z  \\
\hline
0      &        0.00  &      0.000     &     0.000    &      0.000   &   0.000  &     0.000    &      0.000    &  0.000  \\
1      &     2192.04  &     -0.009     &    -0.009    &     -0.009   &  -0.009  &    -0.001    &      0.000    &  0.000  \\
2      &     4256.71  &     -0.018     &    -0.016    &     -0.017   &  -0.018  &    -0.001    &     -0.001    &  0.000  \\
3      &     6198.28  &     -0.026     &    -0.024    &     -0.025   &  -0.025  &    -0.002    &     -0.001    &  0.000  \\
4      &     8020.34  &     -0.033     &    -0.030    &     -0.032   &  -0.033  &    -0.003    &     -0.001    &  0.000  \\
5      &     9725.84  &     -0.040     &    -0.036    &     -0.038   &  -0.039  &    -0.003    &     -0.001    &  0.000  \\
6      &    11317.03  &     -0.046     &    -0.042    &     -0.044   &  -0.045  &    -0.003    &     -0.001    &  0.000  \\
7      &    12795.56  &     -0.051     &    -0.047    &     -0.050   &  -0.051  &    -0.004    &     -0.001    &  0.000  \\
8      &    14162.40  &     -0.056     &    -0.052    &     -0.055   &  -0.056  &    -0.004    &     -0.001    &  0.000  \\
9      &    15417.90  &     -0.061     &    -0.056    &     -0.059   &  -0.061  &    -0.004    &     -0.001    &  0.000  \\
10     &    16561.70  &     -0.065     &    -0.060    &     -0.063   &  -0.065  &    -0.005    &     -0.001    &  0.000  \\
11     &    17592.67  &     -0.068     &    -0.063    &     -0.067   &  -0.068  &    -0.005    &     -0.001    &  0.000  \\
12     &    18508.81  &     -0.072     &    -0.066    &     -0.070   &  -0.072  &    -0.005    &     -0.001    &  0.000  \\
13     &    19307.16  &     -0.074     &    -0.069    &     -0.073   &  -0.074  &    -0.005    &     -0.001    &  0.000  \\
14     &    19983.67  &     -0.076     &    -0.071    &     -0.075   &  -0.077  &    -0.005    &     -0.001    &  0.000  \\
15     &    20533.04  &     -0.078     &    -0.073    &     -0.077   &  -0.078  &    -0.005    &     -0.001    &  0.000  \\
16     &    20948.70  &     -0.079     &    -0.074    &     -0.078   &  -0.080  &    -0.005    &     -0.001    &  0.000  \\
17     &    21223.28  &     -0.080     &    -0.075    &     -0.079   &  -0.081  &    -0.005    &     -0.001    &  0.001  \\
18     &    21352.91  &     -0.080     &    -0.075    &     -0.079   &  -0.081  &    -0.005    &     -0.001    &  0.001  \\
19     &    21375.30  &     -0.080     &    -0.075    &     -0.079   &  -0.081  &    -0.005    &     -0.001    &  0.001  \\
  \hline
  \hline
  \end{tabular}
  \mbox{}\\
  \begin{flushleft}
  $^a$ Indicative non-relativististic Born-Oppenheimer values obtained with basis-set-extrapolated a[5,6]z energies; the extrapolation formula is $E_n = E_\infty + A e^{-\alpha \sqrt{n}}$. Reported values have an estimated error of less than $0.10 \times v$~\cm. \\
  $^b$ Taken from ref. \cite{Moss1993}.\\
  $^c$ This work, using the a5z
  \end{flushleft}
  \end{table}

\begin{table}
\caption{QED corrections for \htwo\  for  $J=0$ vibrational levels. \label{table.qed.H2}}
\mbox{}\\
\begin{tabular}{r  r @{\hskip 0.25in} rrr r@{\hskip 0.25in} rrr r@{\hskip 0.2in}r }
\hline
\hline
       &&  \multicolumn{3}{c}{exact$^b$}                                 && \multicolumn{3}{c}{this work}   &        &\multicolumn{1}{c}{error$^c$}   \\
\cline{3-5}\cline{7-9}
  $v$  &        BO$^a$ &   $\alpha^2$  &  $\alpha^3$&         total &&  $\alpha^2$  &  $\alpha^3$   & total     &&     total  \\
\hline
0      &        0.00   &         0.00  &       0.00   &       0.00  &&      0.00    &      0.00     &   0.00    &&      -0.00   \\
1      &    4,163.40   &         0.02  &       -0.02  &       0.00  &&      0.03    &     -0.02     &   0.01    &&      -0.01   \\
2      &    8,091.16   &         0.04  &       -0.04  &       0.00  &&      0.06    &     -0.04     &   0.01    &&      -0.01   \\
3      &   11,788.14   &         0.05  &       -0.06  &      -0.01  &&      0.07    &     -0.06     &   0.01    &&      -0.02  \\
4      &   15,257.39   &         0.06  &       -0.08  &      -0.02  &&      0.08    &     -0.08     &   0.00    &&      -0.02  \\
5      &   18,499.88   &         0.06  &       -0.09  &      -0.03  &&      0.09    &     -0.10     &  -0.02    &&      -0.02 \\
6      &   21,514.30   &         0.05  &       -0.11  &      -0.05  &&      0.08    &     -0.12     &  -0.04    &&      -0.02 \\
7      &   24,296.64   &         0.04  &       -0.12  &      -0.08  &&      0.07    &     -0.13     &  -0.07    &&      -0.02\\
8      &   26,839.64   &         0.02  &       -0.13  &      -0.12  &&      0.04    &     -0.15     &  -0.10    &&      -0.02\\
9      &   29,131.99   &        -0.02  &       -0.15  &      -0.16  &&      0.01    &     -0.16     &  -0.15    &&      -0.01\\
10     &   31,157.32   &        -0.06  &       -0.16  &      -0.22  &&     -0.04    &     -0.17     &  -0.21    &&      -0.01\\
11     &   32,892.55   &        -0.12  &       -0.17  &      -0.29  &&     -0.10    &     -0.19     &  -0.29    &&      -0.00\\
12     &   34,305.64   &        -0.20  &       -0.18  &      -0.37  &&     -0.18    &     -0.20     &  -0.38    &&       0.01\\
13     &   35,352.20   &        -0.29  &       -0.18  &      -0.48  &&     -0.29    &     -0.21     &  -0.50    &&       0.02\\
14     &   35,970.80   &        -0.42  &       -0.19  &      -0.61  &&     -0.43    &     -0.22     &  -0.65    &&       0.04\\
\hline
\hline
\end{tabular}
\mbox{}\\
\begin{flushleft}
$^a$ Using the very accurate BO potential energy points by Pachucki \cite{Pachucki2010}.\\
$^b$ From Komasa \emph{et al.} \cite{KomasaJCTC2011};  corrections to order $\alpha^4$ were also estimated in ref. \cite{KomasaJCTC2011} but contribute by less than 0.002~\cm{} for all energy levels.\\
$^c$ exact - this work
\end{flushleft}
\end{table}

\begin{table}

 \caption{QED corrections, $\Delta V_{\rm QED}$ in \cm,  computed at the 69 MBB geometries \cite{MBB} using
a Molpro and an a5z FCI wavefunction.. \label{table.qedMBB}}
{\scriptsize{
 \begin{tabular}{cccrccccr}
\hline\hline
na  &   nx  &  ny  &   $\Delta V_{\rm QED}$  && na  &   nx  &  ny  &   $\Delta V_{\rm QED}$ \\
\hline
-4  &   0   &  0   &   0.5588   &&     1   &   -1  &  0   &   -0.0890  \\
-3  &   0   &  0   &   0.3879   &&     1   &   -2  &  0   &   -0.0611  \\
-2  &   0   &  0   &   0.2403   &&     1   &   -3  &  0   &   -0.0126  \\
-1  &   0   &  0   &   0.1121   &&     1   &   -4  &  0   &   0.0594   \\
0   &   0   &  0   &   0.0000   &&     2   &   -1  &  0   &   -0.1749  \\
1   &   0   &  0   &   -0.0981  &&     2   &   -2  &  0   &   -0.1482  \\
2   &   0   &  0   &   -0.1838  &&     2   &   -3  &  0   &   -0.1022  \\
3   &   0   &  0   &   -0.2580  &&     2   &   -4  &  0   &   -0.0345  \\
4   &   0   &  0   &   -0.3217  &&     3   &   -1  &  0   &   -0.2494  \\
5   &   0   &  0   &   -0.3750  &&     3   &   -2  &  0   &   -0.2236  \\
0   &   -1  &  0   &   0.0095   &&     3   &   -3  &  0   &   -0.1797  \\
0   &   -2  &  0   &   0.0391   &&     3   &   -4  &  0   &   -0.1152  \\
0   &   -3  &  0   &   0.0910   &&     4   &   -1  &  0   &   -0.3133  \\
0   &   -4  &  0   &   0.1677   &&     4   &   -2  &  0   &   -0.2883  \\
-1  &   -1  &  0   &   0.1223   &&     4   &   -3  &  0   &   -0.2461  \\
-1  &   -2  &  0   &   0.1542   &&     5   &   -1  &  0   &   -0.3668  \\
-1  &   -3  &  0   &   0.2101   &&     1   &   1   &  0   &   -0.0891  \\
-2  &   -1  &  0   &   0.2515   &&     1   &   2   &  0   &   -0.0624  \\
-2  &   -2  &  0   &   0.2863   &&     1   &   3   &  0   &   -0.0178  \\
-2  &   -3  &  0   &   0.3466   &&     2   &   1   &  0   &   -0.1750  \\
-3  &   -1  &  0   &   0.4002   &&     2   &   2   &  0   &   -0.1485  \\
-3  &   -2  &  0   &   0.4380   &&     2   &   3   &  0   &   -0.1032  \\
-4  &   -1  &  0   &   0.5719   &&     3   &   1   &  0   &   -0.2493  \\
0   &   1   &  0   &   0.0093   &&     3   &   2   &  0   &   -0.2225  \\
0   &   2   &  0   &   0.0366   &&     4   &   1   &  0   &   -0.3129  \\
0   &   3   &  0   &   0.0820   &&     4   &   2   &  0   &   -0.2847  \\
-1  &   1   &  0   &   0.1218   &&     5   &   1   &  0   &   -0.3658  \\
-1  &   2   &  0   &   0.1504   &&     0   &   0   &  2   &   0.0378   \\
-1  &   3   &  0   &   0.1977   &&     -2  &   0   &  2   &   0.2839   \\
-2  &   1   &  0   &   0.2508   &&     -2  &   0   &  3   &   0.3393   \\
-2  &   2   &  0   &   0.2815   &&     0   &   0   &  3   &   0.0864   \\
-2  &   3   &  0   &   0.3319   &&     0   &   0   &  4   &   0.1568   \\
-3  &   1   &  0   &   0.3995   &&     2   &   0   &  2   &   -0.1483  \\
-3  &   2   &  0   &   0.4328   &&     2   &   0   &  3   &   -0.1032  \\
-4  &   1   &  0   &   0.5713   \\
\hline\hline
\end{tabular}
}}  
\end{table}

\section{Rovibrational calculations for \hp\  with the QED surface }
We used the  DVR3D program suite \cite{jt338} to compute ro-vibrational
energy levels using the same parameters employed in previous studies \cite{jt512,jt526};
energy levels are converged with respect to the nuclear motion problem to 0.001~\cm.
Nuclear motion calculations used the new, accurate, global GLH3P PES
of Pavanello \emph{et al.} \cite{jt512}.
This is the most accurate PES available for \hp\ and
includes a non-relativistic BO component computed using explicitly correlated
Gaussian functions \cite{jt512,jt526,ptf09}, an adiabatic Born-Oppenheimer
diagonal correction (BODC) surface \cite{jt512} and a relativistic 
surface \cite{jt512,jt526}.
The BO, adiabatic and relativistic surfaces
are supposed to be accurate to about $10^{-2}$ \cm  \cite{jt512, jt526}.
Here we combine our QED surface with the other surfaces used previously \cite{jt512}.
Calculations were performed  without and with allowance for non-nadiabatic
effects; results are collected in table~\ref{TableH3plusQED}.

\begin{table}
\caption{Vibrational band origins ($J=0$ energy levels) for \hp\ calculated with various models of non-adiabatic effects and with or without QED corrections. All values are in \cm. \label{TableH3plusQED}}
\begin{tabular}{rr @{\hskip 0.25in}rr @{\hskip 0.25in}rr @{\hskip 0.25in}rr}
\hline\hline
\multicolumn{2}{r}{non-ad.$^a$=}    &  nuc      &  nuc      &      PT  &     PT &   Din  & Din    \\
\multicolumn{2}{r}{QED$^b$=}        &    no     &   yes     &      no  &    yes &      no &   yes   \\
$(v_1,v_2^\ell)$  & obs.$^c$        &              \multicolumn{6}{c}{obs.-calc.}                   \\
\cline{3-8}
\hline
 (0,1$^1$)         &  2521.41        &  -0.18    &  -0.14    &     0.11 &   0.16 &   0.01  &   0.05  \\
 (0,2$^2$)         &  4998.04        &  -0.42    &  -0.33    &     0.14 &   0.23 &  -0.03  &   0.05  \\
 (1,1$^1$)         &  5554.06        &  -0.78    &  -0.71    &    -0.14 &  -0.07 &  -0.35  &  -0.28  \\  
 (0,3$^3$)         &  7492.91        &  -0.74    &  -0.61    &     0.13 &   0.26 &  -0.15  &  -0.03  \\  
 (0,4$^2$)         &  9113.08        &  -0.88    &  -0.73    &     0.04 &   0.19 &  -0.26  &  -0.11  \\
 (2,2$^2$)         & 10645.38        &  -1.05    &  -0.95    &     0.06 &   0.20 &  -0.30  &  -0.16  \\
 (0,5$^1$)         & 10862.91        &  -0.85    &  -0.66    &     0.16 &   0.34 &  -0.18  &   0.00  \\
 (3,1$^1$)         & 11323.10        &  -1.27    &  -1.14    &    -0.02 &   0.11 &  -0.41  &  -0.29  \\
 (0,5$^5$)         & 11658.40        &  -1.08    &  -0.90    &     0.09 &   0.27 &  -0.28  &  -0.10  \\
 (2,3$^1$)         & 12303.37        &  -1.15    &  -0.95    &     0.03 &   0.22 &  -0.35  &  -0.16  \\
 (0,6$^2$)         & 12477.38        &  -1.18    &  -0.98    &    -0.02 &   0.18 &  -0.39  &  -0.19  \\
 (0,7$^1$)         & 13702.38        &  -1.33    &  -1.12    &    -0.21 &   0.00 &  -0.62  &  -0.41  \\
 (0,8$^2$)         & 15122.81        &  -1.28    &  -1.06    &     0.16 &   0.38 &  -0.39  &  -0.18  \\
\mbox{}\\
   RMS$^d$         &                 & 0.99      &   0.84    &   0.12    &     0.22 &   0.33 &   0.19\\
\hline
\end{tabular}
\begin{flushleft}
$^a$ Treatment used for non-adiabatic effects. `nuc' indicates nuclear masses were used (i.e., no allowance made for non-adiabatic effects). `PT' indicated the Polyansky-Tennyson model \cite{jt236}  with constant effective rotational and vibrational masses. `Din' is the model by Diniz \emph{et al.} \cite{jt566}\\
$^b$ Indicates whether the QED correction surface was included or not.\\
$^c$ Experimentally-derived energy levels, from Furthenbacher \emph{et al.} \cite{H3pMARVEL1}.\\  
$^d$ Root-mean-square deviation.
\end{flushleft}
\end{table}


Without inclusion of QED effects, the RMS 
deviation obtained for the vibrational
band origins below 16 000 \cm\ is 0.99 \cm{} using nuclear
masses and no allowance for non-adiabatic effects;
inclusion of QED effects results in a reduction of the RMS 
deviation to 0.84 \cm.
The effect of QED is therefore much larger than the desired accuracy of $10^{-2}$ \cm\  for \hp.
The resulting observed $-$ calculated residues can be ascribed almost completely
to non-adiabatic effects.


To further increase the accuracy non-adiabatic effects have to be
taken into account; at the moment this can be done only in an
approximate way, for example using effective rotational and
vibrational masses (PT model \cite{jt236}) or using the more refined
model by Diniz \emph{et al} \cite{jt566}.

To extend the Diniz \emph{et al} model to higher vibrational states
we first calculated $J=0$ energies and wavefunctions, $\Psi_n$, using
nuclear masses. We 
used these wavefunctions and  the mass surface, $m(\underline{R})$,
given by Diniz \emph{et al} to obtain an improved, effective mass, $m_n$, for
each vibrational state $n$ computed as
$m_n = \langle \Psi_n | m(\underline{R}) | \Psi_n \rangle>$. Energies for $J=0$ were then
recalculated for each vibrational state in turn using the improved (constant)
state-dependent mass.


Calculations with a vibrational mass of 1.0007537~u using the PT model results in
a RMS deviation of 0.12~\cm, see table~\ref{TableH3plusQED}.
Inclusion of QED degrades the RMS deviation to 0.22~\cm{} in this model.
On the other hand in the more refined model
of Diniz \etal\ \cite{jt566} for non-adiabatic effects
inclusion of QED effects leads to a reduction of the RMS deviation
from 0.33~\cm{} without QED effects to 0.19~\cm{} when QED is included.

Table \ref{TableH3plusQED} therefore demonstrates that further work is needed
to improve non-adiabatic models as well as that QED corrections are
indispensable to any calculations which include non-adiabatic corrections
in order to approach observed values.

\section{Conclusions}
We  calculated a QED energy correction surface for \hp\ using the approximate method
of Pyykk{\"o} \etal \cite{jt265}.  This method is benchmarked against accurate
QED calculations for H$_2^+$ and \htwo; the comparisons suggest that our
QED surface for \hp\ should provide QED corrections to rotational-vibration
energy levels with an accuracy better than $0.02$~\cm.
The effect of QED on low-lying energy levels is of the order of 0.2~\cm{} and hence
is much larger than the accuracy of $10^{-2}$ \cm\ which has already
been achieved for all components of \ai\ calculations on
\hp\ with the notable exception of non-adiabatic effects.

Inclusion of QED effects leads to \hp\ energy levels
being reproduced with a RMS 
deviation which is reduced from 0.99 \cm\  to 0.84 \cm{} when no allowance is
made for non-adiabatic effects (nuclear masses used for energy levels calculation).
These calculations, which include highly accurate BO, adiabatic, relativistic and QED effects but
no provision for non-adiabatic effects, therefore represent an accurate characterisation of
the value of non-adiabatic effects for each \hp\ level.
Allowance for non-adiabatic effects using the simple model of PT \cite{jt236}
results in a further reduction of this deviation to 0.22~\cm.
Use of the non-adiabatic model of Diniz \etal shows that in this model the use of QED
corrections reduces the errors in the results by almost a factor of two from 0.33~\cm\ to 0.19~\cm.
This demonstrates the necessity of including
QED corrections in  accurate {\it ab initio} treatments of \hp\ rotation-vibration energy levels;  it
opens the way for the development of an accurate non-adiabatic model which could
potentially reach the $10^{-2}$ \cm\  accuracy necessary for the assignment
of Carrington -- Kennedy \cite{CB82} near-dissociation spectrum of \hp\ and its
isotopologues.

\section*{Acknowledgement}
We  thank the Russian Fund for Fundamental Studies,
and ERC Advanced Investigator Project 267219
for supporting  aspects of this project.

\bibliographystyle{apsrev}
\bibliography{journals_phys,h3p,jtj,mp,alexh3p,papers,gen}

\end{document}